\documentstyle[preprint,prl,aps]{revtex}

\begin{document}
\draft
\title{Electronic structure and string tension of stripes in high-$T_{c}$
superconductors }
\author{Marcus Fleck, Eva Pavarini, and Ole Krogh Andersen}
\address{Max-Planck-Institut f\"{u}r Festk\"{o}rperforschung,\\
Heisenbergstrasse 1, D-70569 Stuttgart, Federal Republic of Germany }
\date{\today}
\maketitle

\begin{abstract}
We present a simple, material-specific model Hamiltonian for stripes and use
it to explain the observed differences between the angle-resolved
photoemission spectra of La$_{2-x}$Sr$_{x}$CuO$_{4}$ and Bi$_{2}$Sr$_{2}$CaCu%
$_{2}$O$_{8+x}$, e.g, along the nodal line and around $(3\pi /4,0).$ Next we
compute the string tension for stripes and find it to be smallest in the
materials with the highest observed $T_{c\max }.$
\end{abstract}


Information about the electronic structure of the high-temperature
superconducting cuprates (HTSC) is needed for a detailed understanding of
their thermodynamic and transport properties, and of the superconductivity
itself. Elastic neutron scattering experiments on La$_{2-0.4-1/8}$Nd$_{0.4}$%
Sr$_{1/8}$CuO$_{4}$ have demonstrated that the CuO$_{2}$ layers are unstable
against the formation of one-dimensional (1D) ``charge stripes'' which act
as domain walls separating $\pi $-shifted antiferromagnetic regions \cite
{Tra95}. These findings have triggered an intense research on models in
which the spin and charge degrees of freedom of the electrons are viewed as
inhomogeneous in real space \cite{Zaa89,Eme93,Ore00}.

The electronic structure of 1D domain walls in the two-dimensional (2D) CuO$%
_{2}$ layers of HTSCs has recently attracted a lot of interest in connection
with angle resolved photoemission spectroscopy (ARPES) on La$_{2-x}$Sr$_{x}$%
CuO$_{4}$ (LSCO) \cite{Ino99}, La$_{2-x-y}$Nd$_{y}$Sr$_{x}$CuO$_{4}$
(Nd-LSCO) \cite{Zho99,Zho00}, and Bi$_{2}$Sr$_{2}$CaCu$_{2}$O$_{8+x}$
(BSCCO) \cite{Din96,Sai98,Chu99,Fen99}. A characteristic feature of the
measured Fermi surfaces (FSs) in the 2D Brillouin zone (BZ) is that they
possess straight segments running parallel to the $(0,0)$-$(\pi ,0)$ and $%
\left( 0,0\right) $-$\left( 0,\pi \right) $ directions \cite{Zho99,Fen99}.
Such fingerprints of a 1D band structure can be explained successfully
within the half-filled stripe model \cite{Sal96,Zaa99,Fle00,Mar00,Wro00}.
According to this model, the FS is a straight line, $\left| k_{y}\right| =%
\frac{\pi }{4},$ in the case of stripes running in the $y$-direction. The
stripe direction, $x$- or $y$-, alternates from one CuO$_{2}$-layer to the
next and between domains within the same layer. However, the ARPES
observation of a strong dependence on the momentum $k_{x}$ perpendicular to
the stripe, in particular the 0.$2$ eV-dispersion along $k_{y}\approx 0$ for 
$\frac{\pi }{2}<k_{x}\leq \pi $ not only in Nd-LSCO \cite{Zho99,Zho00}, but
also in underdoped LSCO\cite{Ino99,Zho00}, and in BSCCO \cite{Din96,Sai98}
even for $0<k_{x}\leq \pi ,$ poses a serious problem for the model. In
addition, the observations of nodal quasi-particles (with momentum along the
diagonal of the 2D BZ) in optimally doped LSCO \cite{Zho00} and BSCCO \cite
{Fen99}, but not in underdoped LSCO \cite{Ino99} and Nd-LSCO \cite{Zho99},
are in direct conflict with the simple stripe picture \cite{Ore00,Zaa99}.
Thus, 1D confinement of spectral weight appears {\em not} to be a generic
feature of the stripe phase.

In this Letter we present the electronic structure of 1D domain walls in a
CuO$_{2}$ layer, calculated with a material-specific model Hamiltonian. The
resulting band structure for an electron moving in an effective
stripe-potential provides a simple explanation for the observed differences
in the ARPES data between Nd-LSCO and BSCCO \cite{Chu99,Fen99}. In addition,
we shall calculate kink-excitation energies and see that they tend to zero
for materials with higher observed $T_{c}$ at optimal doping.

We start from the 2D Hubbard model: 
\[
H=\sum_{{\bf k},\sigma }\varepsilon _{{\bf k}}a_{{\bf k}\sigma }^{\dagger
}a_{{\bf k}\sigma }+U\sum_{{\bf R}}n_{{\bf R\uparrow }}n_{{\bf R\downarrow }%
},
\]
where $\varepsilon _{{\bf k}}$ is the bare conduction band, which is
periodic with primitive translation vectors ${\bf G}_{1}=\left( 2\pi
,0\right) $ and ${\bf G}_{2}=\left( 0,2\pi \right) ,$ and $a_{{\bf k}\sigma }
$ $\left[ a_{{\bf R}\sigma }\right] $ annihilates an electron with momentum $%
{\bf k}$ [at site ${\bf R}=\left( l,m\right) $] and spin $\sigma .$ This
model, with $\varepsilon _{{\bf k}}=-2t\left( \cos k_{x}+\cos k_{y}\right) $
and $U=12t,$ has recently been solved numerically using the dynamical
mean-field (DMF) approximation and a supercell approach\cite{Fle00}. For
doping levels $0.05<x<0.2,$ $y$-oriented half-filled stripes were found to
be stable. Such a band with only nearest-neighbor hopping $\left( t\right) $
is a reasonable approximation for LSCO, but for materials with higher $%
T_{c\max }$ such as BSCCO, further hopping integrals $\left( t^{\prime
},\,t^{\prime \prime },...\right) $ are needed to describe both the FSs
observed with ARPES in overdoped samples and also the energy bands
calculated with the local density approximation (LDA)\cite{And95,Pavarini00}%
. In the present letter, we study the effect of using such material-specific
band shapes. The self-energy in the DMF approximation, $\Sigma _{{\bf R}%
\sigma }\left( \omega \right) ,$ defined in terms of the one-electron Green
function by: 
\[
G_{{\bf RR}^{\prime }\sigma }^{-1}\left( \omega \right) =\left[ \omega +\mu
-\Sigma _{{\bf R}\sigma }\left( \omega \right) \right] \delta _{{\bf RR}%
^{\prime }}-\sum_{{\bf k}}\varepsilon _{{\bf k}}e^{i{\bf k\cdot }\left( {\bf %
R-R}^{\prime }\right) },
\]
was found in Ref. \cite{Fle00} to have a spin, site, and energy dependence
which, for low energies and for the doping levels $x\equiv 1/\left(
4N\right) =1/8$ and 1/12, we now model by: 
\begin{eqnarray}
\Sigma _{{\bf R}\sigma }\left( \omega \right)  &=&-\frac{1}{2}\sigma V\sin
\left( {\bf Q\cdot R}\right) -\omega \lambda -i\delta ,\quad \left| \omega
\right| <1.5t,  \nonumber \\[0.15cm]
{\bf Q} &=&\pi \left( 1-\frac{1}{2N},1\right) =\left( 2N-1\right) {\bf g}%
_{1}+N{\bf g}_{2}.  \label{V}
\end{eqnarray}
This yields a low-energy electronic structure which is the bare band, $%
\varepsilon _{{\bf k}},$ subject to a magnetic stripe potential of width $V$
and a mass renormalization $1+\lambda .$ In the $t$-$J$ model, the latter is
due to the formation of spin polarons. From the DMF calculations we obtain: $%
V=3t$ and $\lambda =2,$ for $x=1/8.$ The spin-periodicity of the stripe
potential is $4N$ so that the primitive translation vectors are ${\bf T}%
_{1}=\left( 4N,0\right) $ and ${\bf T}_{2}=\left( 2N,1\right) $ in real
space, and ${\bf g}_{1}=\pi \left( \frac{1}{2N},-1\right) $ and ${\bf g}%
_{2}=\pi \left( 0,2\right) $ in reciprocal space. The observed\cite{Tra95}
magnetic (charge) Bragg peak is ${\bf Q}$ $\left( -2{\bf Q}\right) .$ The
values of $\sin \left[ {\bf Q}\cdot \left( l,0\right) \right] $ for $l$
running from 0 to respectively 7 and 11 are: $0,\,\sqrt{2}/2,\,-1,\,\sqrt{2}%
/2,\,0,\,-\sqrt{2}/2,\,1,\,-\sqrt{2}/2$ and $0,\,1/2,\,-\sqrt{3}%
/2,\,1,\,...\;.$ In the limit when $V\gg t,$ the approximate form (\ref{V})
of the self-energy yields a $1D$ band with energy $\sim \left[ \pm \left(
V/2\right) \sin \left\{ {\bf Q}\cdot \left( l,0\right) \right\} -\mu \right]
/\left( 1+\lambda \right) $ for each row of atoms. Here and in the
following, the upper (lower) sign is for a spin-up (down) electron. The
zero-energy, spin-degenerate, metallic stripe bands centered at $l$=0 and $2N
$ have the dispersions $\varepsilon _{\left( k_{x},k_{y}\right) =\left( \pi
/2,k_{y}\right) }/\left( 1+\lambda \right) $ and $\varepsilon _{\left( \pi
/2,k_{y}+\pi \right) }/\left( 1+\lambda \right) ,$ respectively.

Since for the interpretation of ARPES spectra we need the $\left| {\bf k,}%
\sigma \right\rangle $-projection and since the stripe-potential takes the
simple form (\ref{V}), which couples $\left| {\bf k,}\sigma \right\rangle $
to just $\left| {\bf k-Q,}\sigma \right\rangle $ and $\left| {\bf k+Q,}%
\sigma \right\rangle $, it is convenient to transform to the $\left| {\bf k}%
+n{\bf Q,\,}\sigma \right\rangle $-representation, with the integer $n$
taking $4N$ consecutive values. As a result: 
\begin{eqnarray*}
&&\left\langle {\bf k+}n{\bf Q,}\sigma \left| G\left( \omega \right) \right| 
{\bf k+}n^{\prime }{\bf Q,}\sigma ^{\prime }\right\rangle = \\
&&\quad \quad \quad \quad \quad \quad \delta _{\sigma \sigma ^{\prime }} 
\left[ \left\{ \omega \left( 1+\lambda \right) +\mu +i\delta \right\} 1-H_{%
{\bf k}}\right] _{nn^{\prime }}^{-1},
\end{eqnarray*}
where the un-renormalized Hamiltonian matrix, 
\[
H_{{\bf k}}\equiv \left[ 
\begin{array}{ccccccc}
\varepsilon _{{\bf k-}\left( 2N-1\right) {\bf Q}} & . & 0 & 0 & 0 & . & \pm 
\frac{V}{4} \\ 
. & . & . & . & . & . & . \\ 
0 & . & \varepsilon _{{\bf k}-{\bf Q}} & \pm \frac{V}{4} & 0 & . & 0 \\ 
0 & . & \pm \frac{V}{4} & \varepsilon _{{\bf k}} & \pm \frac{V}{4} & . & 0
\\ 
0 & . & 0 & \pm \frac{V}{4} & \varepsilon _{{\bf k+Q}} & . & 0 \\ 
. & . & . & . & . & . & . \\ 
\pm \frac{V}{4} & . & 0 & 0 & 0 & . & \varepsilon _{{\bf k+}2N{\bf Q}}
\end{array}
\right] , 
\]
is tri-diagonal and periodic. It is isomorphic with the Hamiltonian for a $%
4N $-atomic ring with on-site energies $\varepsilon _{{\bf k+}n{\bf Q}}$ and
nearest-neighbor hopping integrals $\mp V/4.$

For the bare band we shall use the LDA Cu-O antibonding $pd\sigma $-band as
obtained by integrating out the high-energy degrees of freedom\cite
{And95,Pavarini00}: 
\begin{equation}
\varepsilon _{{\bf k}}=-\left[ (1-p)u_{{\bf k}}+\frac{2rv_{{\bf k}}^{2}}{%
1-2ru_{{\bf k}}}\right] \left/ \dot{d}\right. .  \label{conductionband}
\end{equation}
Here, $u_{{\bf k}}\equiv \frac{1}{2}(\cos k_{x}+\cos k_{y}),$\ $v_{{\bf k}%
}\equiv \frac{1}{2}(\cos k_{x}-\cos k_{y}),$\ $1/\dot{d}$ is the band-width
parameter, and $r$ and $p$ are band-shape parameters. Whereas $1/\dot{d}%
\equiv 2t_{pd}^{2}/\left[ \mu -\left( \varepsilon _{p}+\varepsilon
_{d}\right) /2\right] $ is always about 1.6 eV, $r$ contains the essential
materials dependence and was recently observed to increase with $T_{c\max }$%
\cite{Pavarini00}. It describes the conduction-band content of the
axial-orbital, a hybrid between Cu $4s,$ Cu $3d_{3z^{2}-1},$ apical oxygen $%
2p_{z},$ and farther orbitals. The axial orbital lowers the energy of the
conduction band near $\left( \pi ,0\right) ,$ but not along the nodal line.
The parameter $p$ describes buckling-induced plane-oxygen $2p_{z}$-content
and is non-zero in materials such as YBa$_{2}$Cu$_{3}$O$_{7}$ and BSCCO. For
a flat layer, $p=0,$ whereas for a layer so buckled that $p>\left(
1-2r\right) ^{2},$ the saddlepoint is bifurcated away from $\left( \pi
,0\right) $ \cite{And95}$.$ If we expand 1/$\left( 1-2ru_{{\bf k}}\right) $
in (\ref{conductionband}), $\varepsilon _{{\bf k}}$ takes the form of a
one-band Hamiltonian with $t=\left[ 1-p+o\left( r\right) \right] /4\dot{d},$ 
$t^{\prime }=\left[ r+o\left( r\right) \right] /4\dot{d},$ $t^{\prime \prime
}=t^{\prime }/2+o\left( r\right) ,$ etc.. Hence, $t/\left( 1+\lambda \right)
\approx \left( 1-p\right) \times 125\,$meV is at the order of the exchange
coupling $J,$ and $r$ gives the {\em range} of the hopping. For LSCO, $r\sim
0.1$ and $p=0,$ while for BSCCO and YBCO, $r\sim 0.3$ and $p\sim 0.17\,$\cite
{And95,Pavarini00}.

In Fig. \ref{Aknokink} we show for 1/8 hole-doped LSCO and BSCCO the
calculated bands projected onto $\left| {\bf k}\right\rangle $ in the large
BZ. The spectrum for LSCO agrees well with that obtained from the full DMF
calculation with only nearest-neighbor hopping\cite{Fle00}: The two metallic
stripe bands are separated by gaps from the below-lying three magnetic
valence bands and from the above-lying three magnetic conduction bands. In
the $\left| {\bf k}\right\rangle $-projection, the stripe band centered at $%
l $=0 and with bottom along $\Gamma $X is seen near X, where it is $\sim $75
meV below the chemical potential $\mu .$ Proceeding now in the $y$%
-direction, this stripe band disperses upwards and passes $\mu $ near $(\pi ,%
\frac{\pi }{4}).$ The top of the band is seen again near Y, where it is $%
\sim $75 meV above $\mu .$ From Y towards $\Gamma ,$ the band initially
stays flat and as it starts to disperse downwards through $\mu $, it looses
its $\left| {\bf k}\right\rangle $-character. Finally, from $\Gamma $ along
the nodal direction towards M, the stripe band picks up $\left| {\bf k}%
\right\rangle $-character only when, near $\left( \frac{\pi }{2},\frac{\pi }{%
2}\right) ,$ it is above $\mu ;$ the nodal direction therefore appears to be
gapped.

In BSCCO the stripe bands indirectly overlap the magnetic valence and
conduction bands. This is due to the pushing down by the axial orbital of
the $\varepsilon _{{\bf k}}$ band near X and Y. Along the nodal line,
however, the $\varepsilon _{{\bf k}}$ band is completely --and the $%
\varepsilon _{{\bf k}-{\bf Q}}$ and $\varepsilon _{{\bf k}+{\bf Q}}$ bands
are nearly-- unaffected so that the electronic structure is like in LSCO,
but $\mu $ lies lower because it is dragged down with the bands near X and
Y. As a consequence, $\mu $ straddles off the top of the magnetic valence
band along the nodal line. Hence, if the striped phase of BSCCO exists in
the superconducting state, it will have nodal quasiparticles contributed by
the magnetic valence band. Near X and Y, $\mu $ lies higher with respect to
the bands than in LSCO. It even touches the upper part of the stripe band at
Y, as well as the equivalent part of the other ($l$=4-centered) stripe band
at $\left( \frac{3}{4}\pi ,0\right) .$ This behavior agrees quantitatively
with the ARPES measurements on BSCCO \cite{Chu99,Fen99}. In particular, our
calculation identifies both the extra electronic component observed near $%
\left( \frac{\pi }{2},\frac{\pi }{2}\right) ,$ which arises from the
antiferromagnetic regions separated by 1D domain walls, and also the piece
of FS observed near $\left( \frac{3}{4}\pi ,0\right) ,$ which is part of the 
$l$=4-centered stripe band.

The tri-diagonal form of $H_{{\bf k}},$ together with the smallness of $%
V/8t, $ allows us to discuss these features analytically: As seen in Fig. 
\ref{Aknokink}, the shape of the $\varepsilon _{{\bf k}}$ band is left
fairly intact, except where it is crossed by the $\varepsilon _{{\bf k-Q}}$
or the $\varepsilon _{{\bf k+Q}}$ band, or where it is crossed by the $%
\varepsilon _{{\bf k-}2{\bf Q}}$ or the $\varepsilon _{{\bf k+}2{\bf Q}}$
band {\em and} the corresponding intermediate band, $\varepsilon _{{\bf k-Q}%
} $ or $\varepsilon _{{\bf k+Q}},$ is close by. At such crossing points, $H_{%
{\bf k}} $ may be truncated to a $2\times 2$ or $3\times 3$ matrix.

Along the nodal line $\Gamma $M, the $\varepsilon _{{\bf k}}$ band is
crossed by the $\varepsilon _{{\bf k}+{\bf Q}}$ and $\varepsilon _{{\bf k}-%
{\bf Q}}$ bands, which are nearly parallel and have $\varepsilon _{{\bf k}+%
{\bf Q}}>$ $\varepsilon _{{\bf k}-{\bf Q}}.$ The gap surrounding $\mu $ in
LSCO and lying just above $\mu $ in BSCCO is therefore primarily due to the
interaction with the $\varepsilon _{{\bf k}-{\bf Q}}$ band. The
(un-renormalized) energies are then: $\varepsilon _{+}\pm \left[ \varepsilon
_{-}^{2}+(V/4)^{2}\right] ^{1/2},$ where $\varepsilon _{\pm }\equiv \left(
\varepsilon _{{\bf k}}\pm \varepsilon _{{\bf k-Q}}\right) /2,$ and the
anti-bonding state is part of the $l$=0-centered stripe band while the
bonding state is part of a magnetic valence band.

Next we consider ${\bf k=}\left( \frac{3}{4}\pi ,0\right) .$ Since the XM
line is a mirror, $\varepsilon _{{\bf k}-2{\bf Q}}=\varepsilon _{{\bf k}}$
for all ${\bf k}$ along the line $\left( \frac{2N-1}{2N}\pi ,k_{y}\right) ,$
and $\varepsilon _{{\bf k}-{\bf Q}}\sim $ $\varepsilon _{{\bf k}}$ when $%
k_{y}=0.$ The Hamiltonian is thus like that of an $aba$ molecule: A
non-bonding state, $\left( \left| {\bf k}-2{\bf Q}\right\rangle -\left| {\bf %
k}\right\rangle \right) /\sqrt{2},$ remains at $\varepsilon _{{\bf k}-2{\bf Q%
}}=\varepsilon _{{\bf k}},$ while the energies of the bonding and
anti-bonding states are $\varepsilon _{+}\pm \left[ \varepsilon
_{-}^{2}+2(V/4)^{2}\right] ^{1/2}.$ Whereas the bonding state is part of the
magnetic valence band, the non- and anti-bonding states are part of the
metallic stripe bands centered at respectively $l$=0 and $2N.$

An important scale of energies which contribute spectral weight
perpendicular to the stripes is set by the transverse stripe fluctuations 
\cite{Vie94}. We have calculated the energies of various elementary kink
excitations relevant for the string tension of the stripes by displacing the
stripe potential and using periodic boundary conditions with an $8\times 6$
cell. The smallest energies were found for the excitations shown in the top
panel of Fig. \ref{kinkexcit}. As shown in the bottom panel, the string
tension is calculated to decrease with the range-parameter $r,$ once $%
r\gtrsim 0.25,$ and to vanish for $r\sim 0.4.$ Hence, the energy cost of
creating a kink is compensated by the energy gain due to long-range hopping
which allows for coherent escape of holes sideways to the string. Since $r$
is strongly material dependent\cite{Pavarini00}, so is the string tension.
Also, as may seem obvious from the two spectra in Fig. \ref{Aknokink}, the
energy gained by stripe formation decreases with $r.$ It has been observed
that for a large number of hole-doped HTSCs, $r$ calculated with the LDA
correlates with the measured $T_{c}$ at optimal doping\cite{Pavarini00}.
Specifically, as $r$ increases from $\sim $0.1, to $\sim $0.3, and to $\sim $%
0.4, $T_{c\max }$ increases from $\sim $25\thinspace K, to $\sim $%
60\thinspace K, and to $\sim $135\thinspace K. Needless to say, the reason
for this correlation is unknown, but the present results indicate that
suppression of stripes might increase $T_{c}.$ In addition, our results in
Fig. \ref{kinkexcit} show that once $r\gtrsim 0.2,$ the string tension is
softened by dimpling or buckling of the CuO$_{2}$ planes as it occurs in
bilayered YBa$_{2}$Cu$_{3}$O$_{7}$ and BSCCO. Finally, we have found that
the domain-wall geometry with the shorter perpendicular distance has the
lower energy. This implies that domain-wall dynamics does weaken the
potential barrier for pair-tunneling between charge stripes\cite{Kiv98}.

In Figs. \ref{Aklscokink} and \ref{Akybcokink} we show how the kink
distortions influence the single-particle spectra calculated for LSCO and
BSCCO, respectively. The gaps, say at $\left( \frac{2N-1}{2N}\pi ,0\right) $
between the two stripe bands and between the lower stripe band and the upper
magnetic valence band, tend to get filled. Even for LSCO, the parallel
excursion of neighboring domain walls shown in the upper left of Fig. \ref
{kinkexcit} nearly fills the charge gap and gives strong spectral intensity
in the nodal direction. Our LSCO spectra shown at the bottom of Fig. \ref
{Aklscokink} are in accord with the experimental observation of FS features
near $\left( \frac{\pi }{2},\frac{\pi }{2}\right) $ generated by dynamic
stripes in La$_{1.85}$Sr$_{0.15}$CuO$_{4}$ \cite{Ino99,Zho00}. One should,
however, note the difference to BSCCO where nodal spectral intensity from
the magnetic valence band exists in already the ordered stripe phase. Kinks
thus influence the spectrum in the nodal direction and near $\mu $ far less
for BSCCO than for LSCO. Consistent with our energy analysis we understand
that transverse stripe fluctuations are weaker in LSCO due to the energy
cost of loosing the charge gap along the nodal direction. Figs. \ref
{Aklscokink} and \ref{Akybcokink} show that dominant FS features, such as
the straight, $k_{x}$-directed stripe-band segments, the magnetic
valence-band sheet near the middle of the nodal line in BSCCO, and the sheet
near $\left( \frac{2N-1}{2N}\pi ,0\right) $ from the $2N$-centered stripe
band in BSCCO, are fairly robust. This finding for the $\left( \frac{2N-1}{2N%
}\pi ,0\right) $-sheet is in accord with recent experimental observations 
\cite{Chu99,Fen99} and clarifies the questions raised in the experimental
papers about the origin of such quasi-particle states.

In conclusion, we have shown that the low-energy spectral properties of
striped phases can be studied with a simple, material-specific model for
electrons moving in an effective stripe potential. This model has allowed us
to understand experimental features, such as: (i) the absence of spectral
weight in the nodal direction for underdoped LSCO \cite{Ino99} and 1/8-doped
Nd-LSCO \cite{Zho99}, (ii) the presence of spectral weight in the nodal
direction for BSCCO \cite{Din96,Sai98,Chu99,Fen99}, (iii) the low-energy
photoemission states around $(\frac{3}{4}\pi ,0)$ in BSCCO \cite{Chu99,Fen99}%
, (iv) the filling of the gap along the nodal direction in the dynamic
striped phase of optimally doped LSCO \cite{Zho99}, and (v) the enhancement
of superconducting fluctuations due to stripe dynamics. Finally, we have
seen that the string tension of a stripe tends to be smallest in the HTSC
materials with the highest observed $T_{c\max }.$

We thank C. Bernhard, O. Gunnarsson, P. Horsch, O. Jepsen, and J. Zaanen for
stimulating discussions.

%
\begin{figure}[tbp]
\caption{Calculated single-particle spectra, $A({\bf k},\protect\omega )$
for ${\bf k}$ along the symmetry directions of the large BZ, of the $y$%
-oriented striped phase for 1/8 doped LSCO (top) and BSCCO (bottom). $%
\Gamma=\left( 0,0\right) ,$ X$=\left( \protect\pi ,0\right) ,$ Y$=\left( 0,%
\protect\pi \right) ,$ M$=\left( \protect\pi ,\protect\pi\right) ,$ S$%
=\left( \protect\pi /2,\protect\pi /2\right) $.}
\label{Aknokink}
\end{figure}

\begin{figure}[tbp]
\caption{Calculated energy cost for creating localized kinks in an $8 \times
6$ cell $(l=-2,.....,5$ and $m=0,...,5)$ for 1/8 hole doping as a function
of the range parameter $r$. Circles are for flat layers ($p=0$) and squares
are for $p=0.17$. As illustrated in the upper panel, the open (closed)
symbols refer to kinks which do (not) conserve the distance between domain
walls.}
\label{kinkexcit}
\end{figure}

\begin{figure}[tbp]
\caption{Calculated spectra of the $y$-oriented striped phase with kinks in
1/8 doped LSCO. The top and the lower-right (left) parts are for the kinks
shown in the upper-right (left) part of Fig. \ref{kinkexcit}. The figures of
the lower panel show the spectral weight in the large BZ, integrated over an
energy window extending from $\protect\mu $ to 25 meV below.}
\label{Aklscokink}
\end{figure}

\begin{figure}[tbp]
\caption{Spectra as in Fig. \ref{Aklscokink} but for BSCCO.}
\label{Akybcokink}
\end{figure}

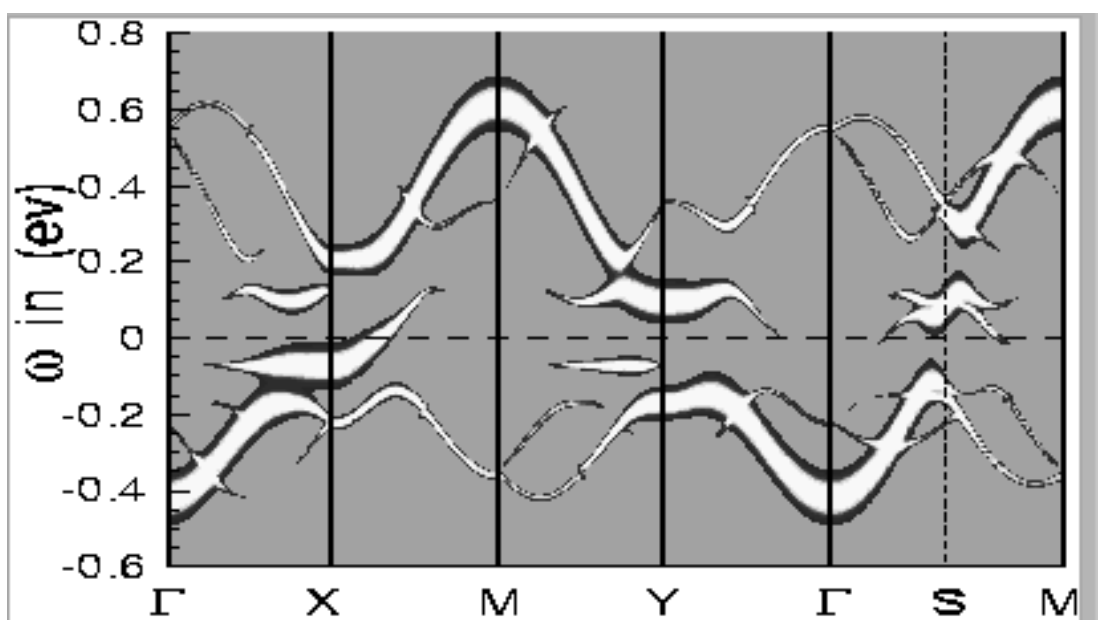
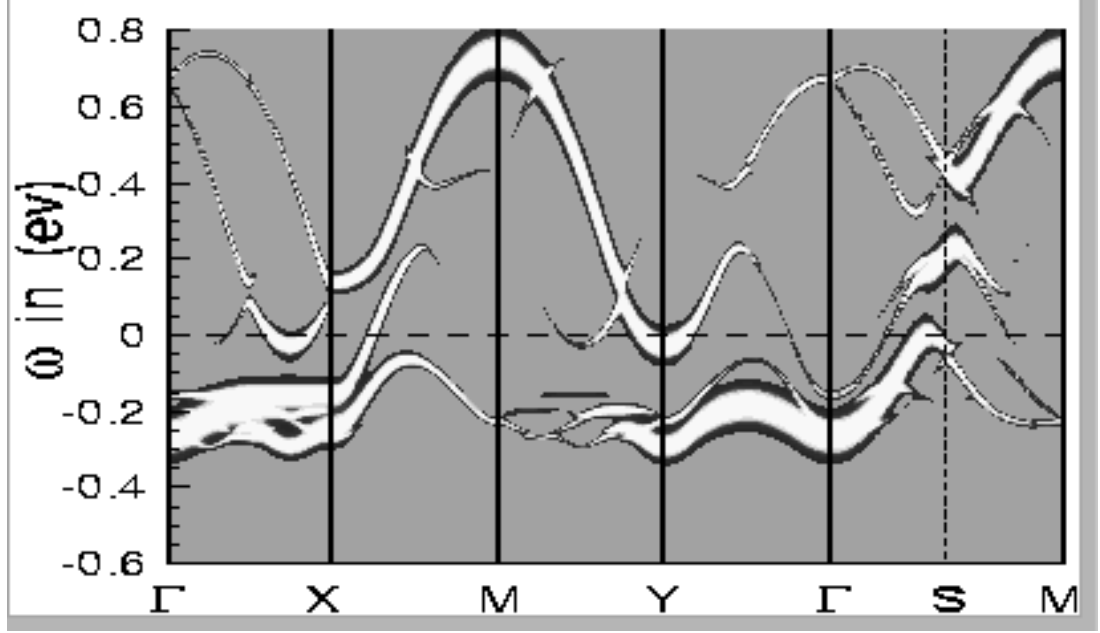
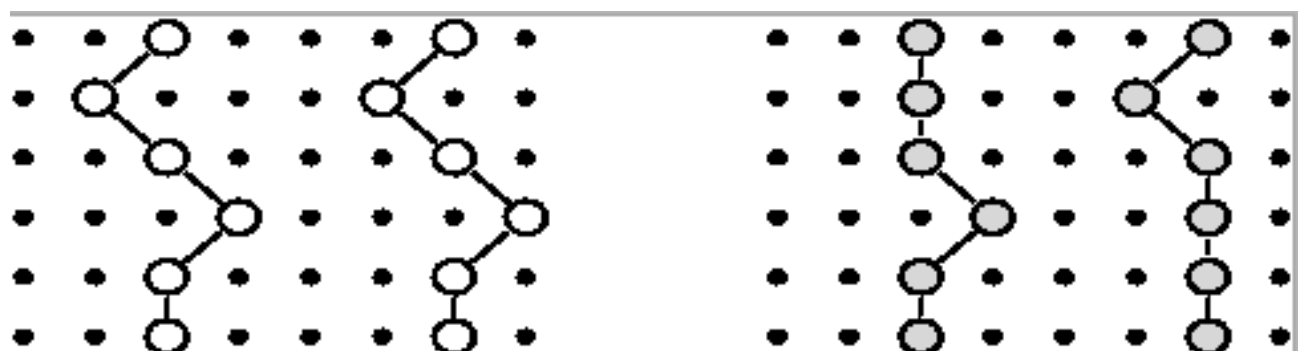
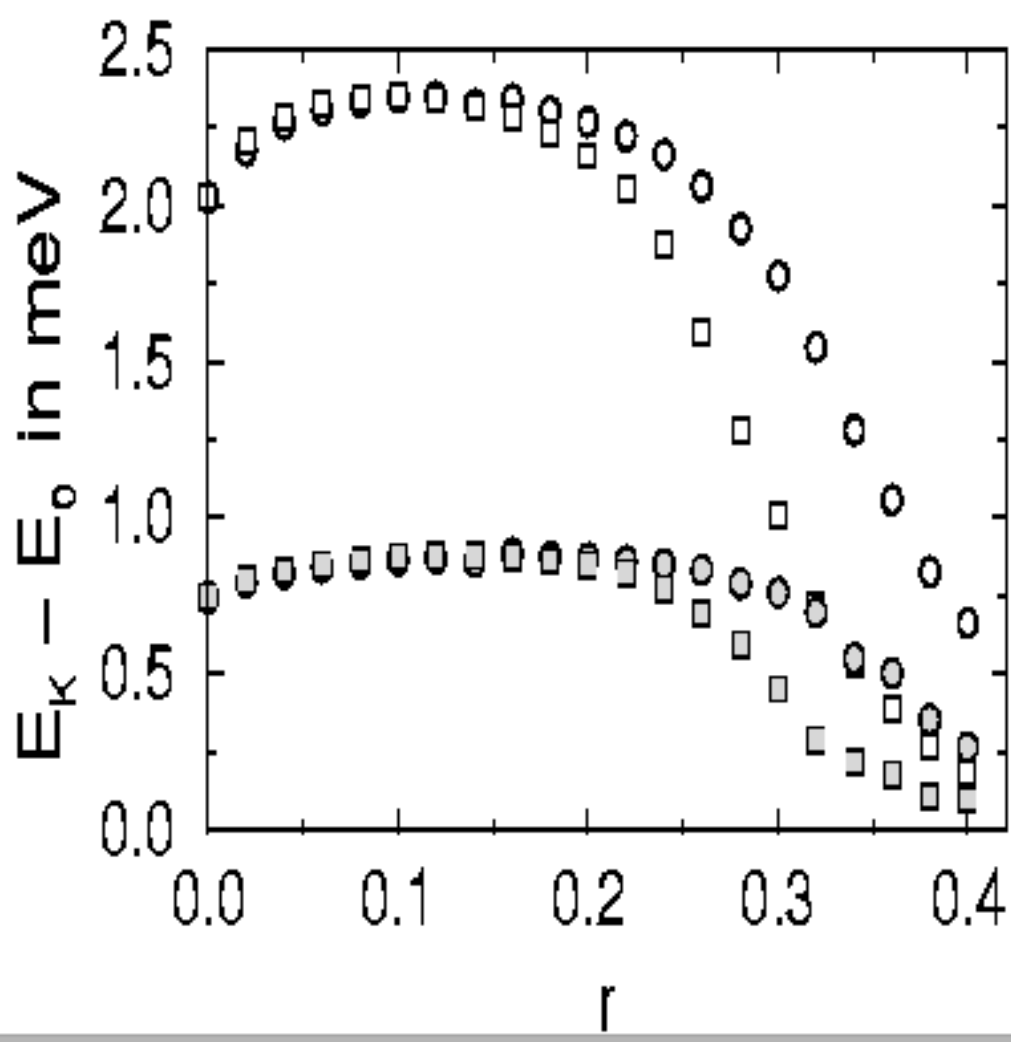
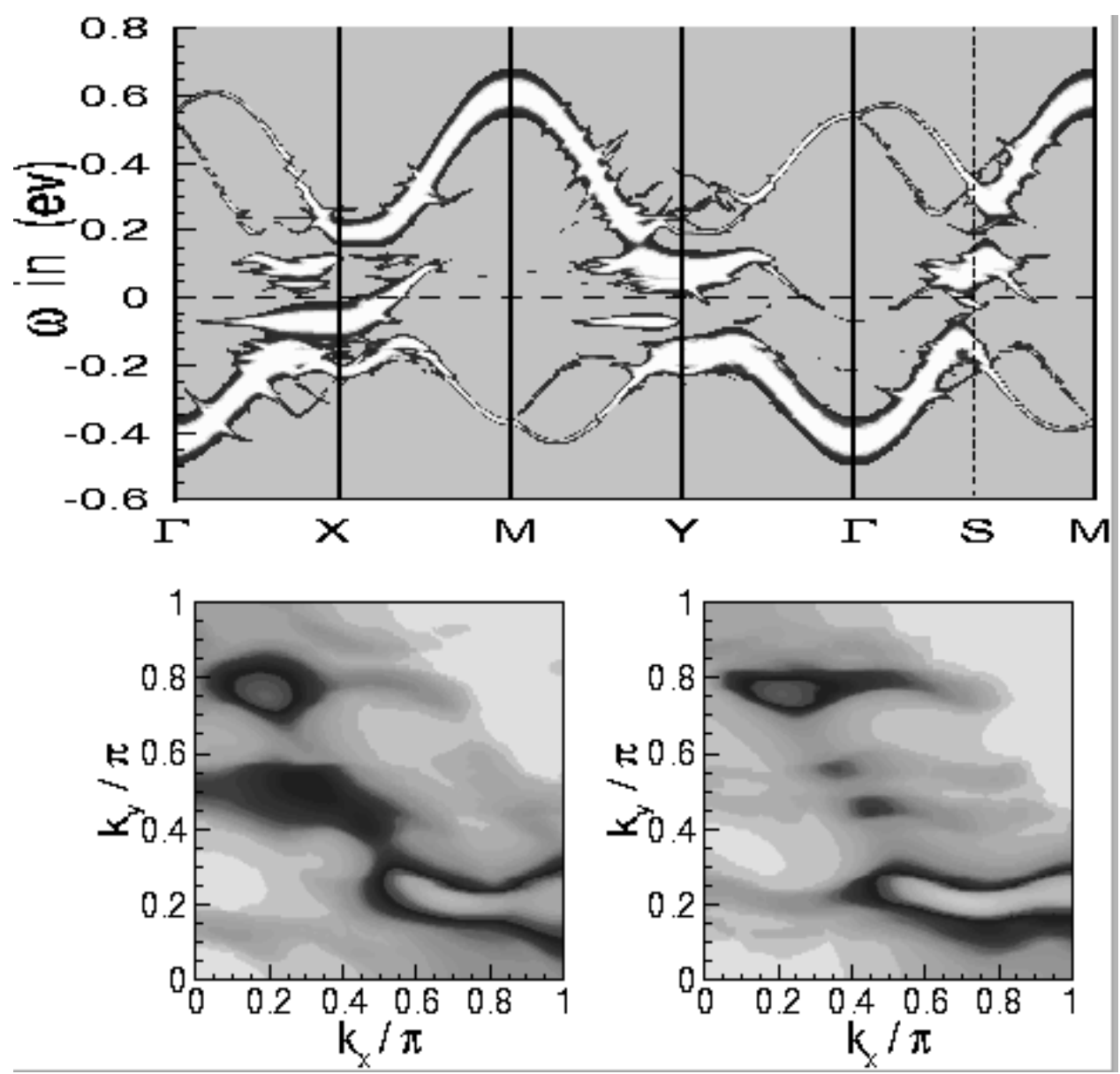
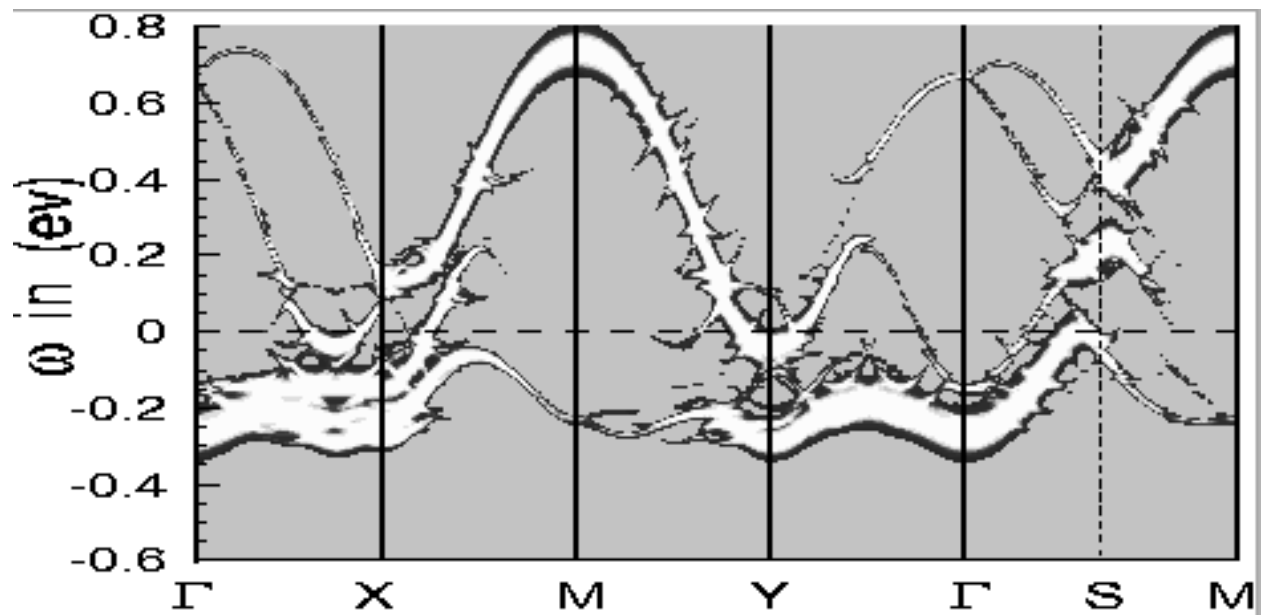
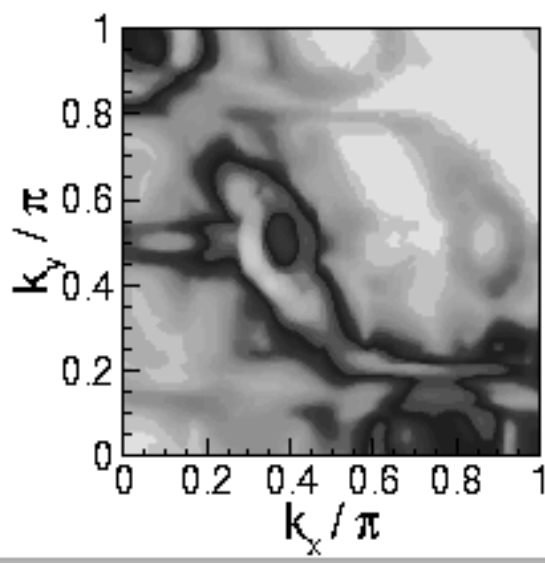
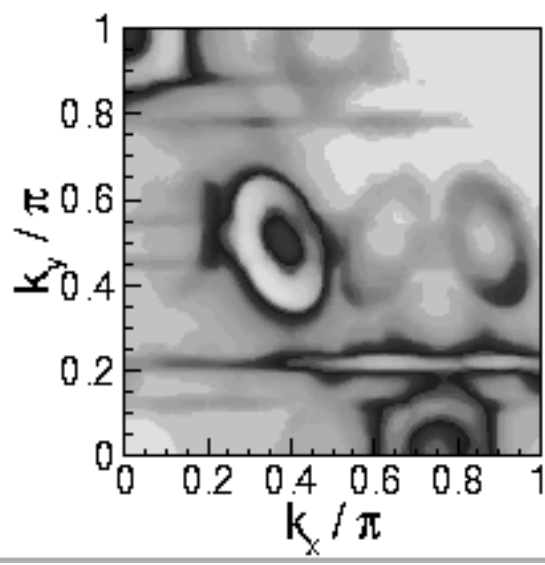
\end{document}